\documentclass[a4paper,11pt]{article}
\usepackage[british]{babel}

\usepackage[includeheadfoot, width=432pt, height=720pt]{geometry}
\usepackage[utf8]{inputenc}
\usepackage{hyperref}
\hypersetup{colorlinks=false, linkcolor=red, citecolor=green, urlcolor=blue-purple}
 \usepackage[splitrule]{footmisc}

\usepackage{verbatim}
\usepackage{lmodern}
\usepackage{lipsum}
\usepackage{booktabs}
\usepackage{caption}
\usepackage{cite}
\usepackage{soul,color,xcolor}

\usepackage{amsthm, amssymb,amsfonts}
\usepackage[tbtags]{amsmath}
\usepackage{bm}
\usepackage{mathtools}
\usepackage{amstext}
\usepackage{braket}
\usepackage{multirow}
\usepackage[normalem]{ulem}

\begin{document}

\def\be{\begin{eqnarray}}
\def\ee{\end{eqnarray}}
\def\ba{\begin{aligned}}
\def\ea{\end{aligned}}
\def\rmd{\mathrm{d}}
\def\ls{\left[}
\def\rs{\right]}
\def\lc{\left\{}
\def\rc{\right\}}
\def\p{\partial}
\def\S{\Sigma}
\def\s{\sigma}
\def\O{\Omega}
\def\a{\alpha}
\def\b{\beta}
\def\g{\gamma}
\def\ad{{\dot \alpha}}
\def\bd{{\dot \beta}}
\def\gd{{\dot \gamma}}
\newcommand{\ft}[2]{{\textstyle\frac{#1}{#2}}}
\def\ib{{\overline \imath}}
\def\jb{{\overline \jmath}}
\def\Re{\mathop{\rm Re}\nolimits}
\def\Im{\mathop{\rm Im}\nolimits}
\def\trace{\mathop{\rm Tr}\nolimits}
\def\rmi{{ i}}
\def\N{\mathcal{N}}
\def\Mp{M_{\rm P}}

\newcommand{\SU}{\mathop{\rm SU}}
\newcommand{\SO}{\mathop{\rm SO}}
\newcommand{\U}{\mathop{\rm {}U}}
\newcommand{\USp}{\mathop{\rm {}USp}}
\newcommand{\OSp}{\mathop{\rm {}OSp}}
\newcommand{\Symp}{\mathop{\rm {}Sp}}
\newcommand{\Sl}{\mathop{\rm {}S}\ell }
\newcommand{\Gl}{\mathop{\rm {}G}\ell }
\newcommand{\Spin}{\mathop{\rm {}Spin}}

\def\hc{c.c.} 
\def\Mp{M_{\text{P}}}
\def\QG{\Lambda_{\text{QG}}}

\definecolor{purple}{rgb}{0.4 ,0, 0.85}
\definecolor{blue-purple}{rgb}{0.18, 0, 0.9}

\newcommand{\MS}[1]{\textcolor{magenta}{[MS: #1]}}
\newcommand{\DL}[1]{\textcolor{magenta}{[DL: #1]}}

\allowdisplaybreaks

\allowbreak

\setcounter{tocdepth}{2}

%%%%%%%%%%%%%%%

\begin{titlepage}
	\thispagestyle{empty}
	\begin{flushright}
		\hfill{MPP-2023-5} \\
		\hfill{LMU-ASC 01/23}		
	\end{flushright}
\vspace{5pt}

	\begin{center}
	      
	     { \Large\bf{
\mbox{\hspace{-16pt} The Scale of Supersymmetry Breaking and the Dark Dimension}}}

		\vspace{25pt}

		{\mbox{\hspace{-15pt}Luis A. Anchordoqui$^{1,2,3}$, Ignatios Antoniadis$^{4,5}$, Niccol\`o Cribiori$^{6}$, Dieter L\"ust$^{6,7}$, Marco Scalisi$^{6}$}}

		\vspace{30pt}

		{
            $^1${\it Department of Physics and Astronomy,\\
Lehman College, City University of New York, NY 10468, USA}
            
            	\vspace{6pt}
            $^2$\mbox{\it Department of Physics, Graduate Center, City University of New York, NY 10016, USA}
            
            	\vspace{6pt}
            $^3${\it Department of Astrophysics, American Museum of Natural History, NY 10024, USA }
            
\vspace{6pt}
            $^4${\it Laboratoire de Physique Th\'eorique et Hautes Energies - LPTHE,\\
Sorbonne Universit\'e, CNRS, 4 Place Jussieu, 75005 Paris, France}       
\vspace{6pt}

$^5${\it Department of Physics, Harvard University, Cambridge, MA 02138, USA}

\vspace{6pt}
            $^6${\it Max-Planck-Institut f\"ur Physik (Werner-Heisenberg-Institut),\\ F\"ohringer Ring 6, 80805, M\"unchen, Germany }     
            
            \vspace{6pt}
            $^7${\it Arnold Sommerfeld Center for Theoretical Physics,\\
Ludwig-Maximilians-Universit\"at M\"unchen, 80333 M\"unchen, Germany}

}
		\vspace{55pt}

		{ABSTRACT}
	\end{center}

	\vspace{0pt}
\noindent
We argue for a relation between the supersymmetry breaking scale and the measured value of the dark energy density $\Lambda$. We derive it by combining two quantum gravity consistency swampland constraints, which tie the dark energy density $\Lambda$ and the gravitino mass $M_{3/2}$, respectively,  
to the mass scale of a light Kaluza-Klein tower and, therefore, to the UV cut-off of the effective theory.
Whereas the constraint on $\Lambda$ has recently led to the Dark Dimension scenario, with a prediction of a single mesoscopic extra dimension of the micron size, we use the constraint on $M_{3/2}$ to infer the implications of such a scenario for the scale of supersymmetry breaking. 
We find that a natural scale for supersymmetry signatures is
\begin{equation}
M={\cal O}\left(\Lambda^{\frac18}\right)={\cal O}({\rm TeV})\, .\nonumber
\end{equation}
This mass scale is  
within reach of LHC and of the next generation of hadron colliders. 
Finally, we discuss possible string theory and effective supergravity realizations of the Dark Dimension scenario with broken supersymmetry.

\bigskip

\end{titlepage}

%%%%%%%%%%%%%%%

%\tableofcontents

%\newpage

%%%%%%%%%%%%%%%%%%%%%

\section{Introduction}
\label{sec:Intro}

At present, there are two main hierarchy problems in theoretical physics. 
The first concerns the value and radiative stability of the electroweak hierarchy $M_{\rm EW}/M_{\rm P} \simeq 10^{-16}$, where $M_{\rm EW}$ is the electroweak symmetry breaking scale and $\Mp$ is the reduced Planck mass. 
The second concerns the value and radiative stability of the cosmological hierarchy \mbox{$\Lambda/M_{\rm P}^4 \simeq 10^{-120}$}, where $\Lambda$ is the dark energy density as measured today \cite{Planck:2018vyg,SupernovaCosmologyProject:1998vns,SupernovaSearchTeam:1998fmf}.

Supersymmetry (SUSY) has been proposed as a field-theoretic scenario in which the first problem can possibly find a solution. 
It is assumed  that such a symmetry be restored at high energies, since no supersymmetric particles have been seen to date. 
The scale $M_{\rm SUSY}$ at which supersymmetry is expected to be broken is unknown and object of intense experimental effort. 
To address the second problem, a novel paradigm has been recently put forward.
It has been named Dark Dimension scenario \cite{Montero:2022prj}, for it predicts the existence of one mesoscopic extra dimension at the micron scale.
This is in fact derived by combining experimental observations and Swampland criteria \cite{Vafa:2005ui,Palti:2019pca,vanBeest:2021lhn,Agmon:2022thq}, which aim to capture universal properties of quantum gravity. They go beyond purely field-theoretic methods by connecting low-energy physics predictions to the ultraviolet quantum gravity cut-off of the effective field theory.
In particular, in the Dark Dimension proposal, the Anti-de Sitter Distance Conjecture (ADC) \cite{Lust:2019zwm} plays a central role so that the length of the Dark Dimension is set by the dark energy scale as $\Lambda^{-\frac14}$, modulo a correction factor \cite{Montero:2022prj,Anchordoqui:2022ejw}.
Despite its recent appearance, the Dark Dimension scenario has already received much attention in the literature \cite{Anchordoqui:2022ejw,Anchordoqui:2022txe,Blumenhagen:2022zzw,Gonzalo:2022jac,Anchordoqui:2022tgp,Anchordoqui:2022svl}, with interesting implications for dark matter physics \cite{Anchordoqui:2022txe,Gonzalo:2022jac,Anchordoqui:2022tgp}.

The purpose of the present paper is to study the connection between these two proposals. As a result, we are led to a prediction on the supersymmetry breaking scale $M_{\rm SUSY}$ from the measured value of the dark energy density $\Lambda$. 
Concretely, we argue for a relation of the type
\begin{equation}
\label{Msusyn=1}
M_{\rm SUSY} \simeq \left(\frac{\Lambda}{\Mp^4}\right)^\frac{1}{8n} \Mp \, ,
\vspace{0.3cm}\end{equation} 
where $n$ is integer or half-integer satisfying $1\leq n \leq 2$. The case with $n=1$ represents the simplest scenario and points at a supersymmetry breaking scale of order\footnote{It was already noted in \cite{Banks:2000fe} that a relation such as $M_{\rm SUSY}\sim\Lambda^{1/8}$ would be interesting for phenomenology. However, in the same work, no derivation is provided for such a formula.}
\begin{equation}
 M_{\rm SUSY}= \mathcal{O}(1- 10)\left(\frac{\Lambda}{\Mp^4}\right)^\frac{1}{8}  \Mp \simeq \mathcal{O}(10- 100) \ {\rm TeV}\,.
\end{equation}
This is intriguingly close to the current bound provided by the Large Hadron Collider (LHC) \cite{ParticleDataGroup:2022pth}. Scenarios with $n>1$ lead to higher supersymmetry breaking scale but still to a gravitino mass $M_{3/2}=\mathcal{O}({\rm GeV - TeV})$.
Therefore, consistency with quantum gravity not only provides a framework to address the smallness of $\Lambda$, but also ties this scale to $M_{\rm SUSY}$ with a predicted value, which might be tested even by current technology.

To reach such a result, we employ the following strategy. 
First, we observe that in an approximately flat background, one can estimate the scale of supersymmetry breaking from the mass $M_{3/2}$ of the gravitino, the superpartner of the graviton, as $M_{\rm SUSY} \simeq \sqrt{M_{3/2}\, \Mp}$.
This assumes that supersymmetry breaking effects, which could be due to expectation values of F- and/or \mbox{D-terms} in an $\mathcal{N}=1$ language, are almost entirely balanced by the gravitino mass that provides the genuinely gravitational contribution to the vacuum energy in supersymmetric theories. 
As such, this is a general feature of quasi-flat backgrounds with broken supersymmetry, which is fairly model independent from details of the microscopic theory. 
The implication is that, given the current value of $\Lambda$, one has to deal just with a single scale and can extract information on $M_{\rm SUSY}$ by studying properties of the gravitino.

Our second step is to associate the gravitino mass to the mass scale of an infinite tower of (Kaluza-Klein) states, as it was already proposed in \cite{Antoniadis:1988jn}. 
In the context of the Swampland program, such an idea has been recently formulated as the Gravitino Conjecture \cite{Cribiori:2021gbf,Castellano:2021yye}. 
It relates the gravitino mass to the quantum gravity cut-off of the theory and it implies that the limit in which the former is small leads to a reduction of the latter.\footnote{Prior to \cite{Cribiori:2021gbf,Castellano:2021yye}, a connection between the weak gravity conjecture, massless charged gravitini and the breakdown of the effective description on a de Sitter background has been pointed out in \cite{Cribiori:2020use} and subsequently also in \cite{DallAgata:2021nnr}. For a discussion on the role of fermions in the Swampland program in general, both spin-1/2 and spin-3/2, see \cite{Palti:2020tsy}.} This conjecture is similar in spirit, and it even coincides in specific situations, to the ADC, which instead relates the mass scale of an infinite tower to the value of the dark energy density. In short, in this context, both $M_{3/2}$ (or equivalently $M_{\rm SUSY}$) and $\Lambda$ are related to infinite towers of states, which are external to the effective description and then provide a connection to the scale of extra-dimensions.

As third step, we distinguish and study two situations. In the first case, the gravitino mass and the dark energy density are related to the same tower of states. This is arguably the simplest scenario, which leads directly to \eqref{Msusyn=1}. % or to a gravitino mass $M_{3/2}=\mathcal{O}({\rm GeV - TeV})$.
In the second case, $M_{3/2}$ and $\Lambda$ are related to different towers. This scenario requires a decoupling of the gravitino mass from the dark energy density and is thus more difficult to realize in concrete models (see, for example, \cite{Coudarchet:2021qwc} for a possible embedding in string theory).

The structure of the paper is as follows. In section \ref{sec:GCC}, we review the Dark Dimension scenario, together with the ADC, and the Gravitino Conjecture. In section \ref{sec:Scenarios}, we discuss the two possible scenarios with a single or double Kaluza-Klein tower and extract predictions for the scale of supersymmetry breaking. In section \ref{sec:eftmodels}, we provide possible embeddings into string theory and their effective low energy supergravity description in four dimensions. In section \ref{sec:concl}, we collect our conclusions. Throughout the paper, we work with the reduced Planck mass, $M_{\rm P} =1/{\sqrt{8\pi G_N}}\simeq 2.48 \times 10^{18}$ GeV.

\section{Dark energy, gravitino, and Kaluza-Klein towers}
\label{sec:GCC}

The starting point of the Dark Dimension scenario \cite{Montero:2022prj} is the anti-de Sitter distance conjecture (ADC) \cite{Lust:2019zwm}, which relates the {\it dark energy} density $\Lambda$ to the mass scale $ m_1$ of an infinite tower of states.  
This tower is typically identified with Kaluza-Klein (KK) states or with modes, whose mass is instead governed by the string coupling \cite{Lee:2018urn,Lee:2019wij}.  
In the limit $\Lambda\rightarrow 0$, the ADC implies a decreasing of the quantum gravity cut-off, which in the present work we assume to be given by the species scale \cite{Dvali:2007hz,Dvali:2007wp}. 
In four dimensions, the relation put forward in~\cite{Lust:2019zwm} is 
\begin{equation}\label{masstowerADC}
 m_1\sim\left(\frac{|\Lambda|}{M_{\rm P}^4}\right)^{a}\,M_{\rm P}\,,
\end{equation}
with $a$ being an order-one, positive parameter. In the same work, it is also argued that the conjecture can be extended to backgrounds with positive energy density, like (quasi-)de Sitter solutions.

The proposal of the Dark Dimension~\cite{Montero:2022prj} is to combine experimental bounds on the size of extra spacetime dimensions~\cite{Hannestad:2003yd} with such an extended ADC.
As a result, a unique prediction is made for the existence of one extra dimension of radius $R_1$, associated to a tower of KK modes with $m_1= 1/R_1$.  
Furthermore, as also explained in \cite{Montero:2022prj}, one has a lower bound on $a$ based on swampland arguments and an upper bound required by unitarity \cite{Lust:2019zwm}, in such a way that 
\begin{equation}\label{bounda}
\frac14 \leq a \leq \frac12\, .
\end{equation}

Concretely, in the Dark Dimension scenario it is proposed that the parameter $a$ entering the ADC relation \eqref{masstowerADC} is given by $a=1/4$. 
Then, this leads to the following relation between the dark energy density and the KK scale,
\begin{equation}\label{masstowerDD}
m_1 = \lambda^{-1}\,\Lambda^{\frac14}\,. 
\end{equation}
Here, $\lambda$ is a parameter which is estimated within the range $10^{-4}< \lambda <10^{-1}$ to accommodate limits on short-range deviations from Newton's gravitational inverse-square law~\cite{Kapner:2006si,Lee:2020zjt} and bounds on neutron star heating by the surrounding cloud of trapped KK gravitons~\cite{Hannestad:2003yd}. 
In addition, if neutrino masses originate from right-handed neutrinos propagating in the bulk, then neutrino oscillation data constrain $m_1 > 2.5~{\rm eV}$ at the 99\% CL, implying $\lambda \lesssim 10^{-3}$~\cite{Anchordoqui:2022svl}. 
In the present work, we assume a mild tuning of $\lambda \simeq 10^{-3}$, which in turn gives an explanation of the cutoff in the cosmic ray spectrum~\cite{Anchordoqui:2022ejw}. 
For $\Lambda \simeq 10^{-120} M_{\rm P}^4$, the prediction is that of a Dark Dimension of radius $R_1 \simeq \lambda \times {\rm mm} \sim \mu {\rm m}$.

As of now, no connection between the Dark Dimension and the scale of supersymmetry breaking has been made. 
In order to perform this step, we focus our attention on one specific particle which is believed to be present in the spectrum of any effective theory arising from supergravity and string theory: the {\it gravitino}. 
In the context of the Swampland program, a Gravitino Conjecture has been proposed \cite{Cribiori:2021gbf,Castellano:2021yye}, which relates the gravitino mass $M_{3/2}$ to the mass scale $m_2$ of an infinite tower of states. Such a tower can be in principle different from the previous one, but it is again typically identified with a tower of KK modes of another dimension of radius $R_2$. 
Similarly to the ADC, the Gravitino Conjecture is associated with a reduction of the species scale in the limit $M_{3/2}\to 0$ \cite{Cribiori:2021gbf}.
The relation proposed in~\cite{Cribiori:2021gbf,Castellano:2021yye} has the form
\begin{equation}\label{masstowerGC}
m_2= \lambda_{3/2}^{-1} \ \left(\frac{M_{3/2}}{M_{\rm P}}\right)^{n}\,M_{\rm
  P}\, ,
\end{equation}
with $n$ being an order-one positive parameter and $\lambda_{3/2}$, in analogy to $\lambda$, another proportionality constant.

As a matter of fact, it turns out that in many effective theories, and also in explicit string models, the ADC and the Gravitino Conjecture are related in the sense that $\Lambda$ and $M_{3/2}$ are not independent. Instead, they are coupled to one another in such a way that the respective parameters $a$ and $n$ obey a relation of the form
\begin{equation}
\label{nak}
\frac na=k\, ,
\end{equation}
where $k$ is a constant that depends on the microscopic realization of the conjectures.
The prototype example of this fact are perhaps supersymmetric anti-de
Sitter vacua, where $M_{3/2}$ is related to the anti-de Sitter
cosmological constant in the following simple way:\footnote{The coefficient $1/3$ is valid for minimal supergravity in four  dimensions. However, it is a fact that for supersymmetric anti-de  Sitter vacua in generic dimensions and with a generic number of  preserved supercharges, the cosmological constant is determined by  the mass of the gravitini up to a numerical coefficient.}
\begin{equation}\label{m32SUSYAdS}
M_{\rm P}^2\,M_{3/2}^2=-{\frac{\Lambda}{ 3}}\, .
%M_{\rm P}^2\,m_{3/2}^2=\mathcal{O}(|\Lambda|)\, .
\end{equation}
This immediately implies that $n=2a$, i.e.~$k=2$, and the two towers are expected to be identical to each other.
In more realistic setups with broken supersymmetry, which is actually necessary for a positive cosmological constant, the two parameters $a$ and $n$ can still depend on each other, but 
 different relations between $a$ and $n$ other than $n=2a$ are possible.

\section{Tower scenarios for SUSY breaking scale}
\label{sec:Scenarios}

The Dark Dimension and the Gravitino Conjecture lead immediately to two possible scenarios, depending on the relation between the corresponding towers of external states. A first possibility is that the dark energy density $\Lambda$ and the gravitino mass $M_{3/2}$ are connected to the same KK tower. A second possibility is that the towers are different.
The choice between one of the two options has diverse implications for the supersymmetry breaking scale. In the following, we discuss both of these scenarios.

\subsection{Scenario with a single Kaluza-Klein tower}
\label{sec:1KKtower}

This is the minimal scenario in which all relevant mass scales in the effective theory are governed by a single one. 
In particular, the gravitino mass and the scale of supersymmetry breaking can directly be determined from the dark energy density.

Assuming that $m_2\equiv m_1$, namely we have a single KK tower, the quantum gravity cut-off $\QG$ is determined by $\QG= m_1^{1/3} \Mp^{2/3}$ (see for example \cite{Hebecker:2018vxz,Blumenhagen:2019vgj,Cribiori:2021gbf,Montero:2022prj,Castellano:2022bvr}). This is obtained in four dimensions decompactifying to five at the species scale.\footnote{We recall that the species scale in a decompactification from $4$ to $4+d$ dimensions is given by \mbox{$\QG = m^{\frac{d}{d+2}}M_{\rm P}^{\frac{2}{d+2}}$}~\cite{Montero:2022prj,Castellano:2022bvr}. Throughout we defined the species scale in terms of the reduced Planck mass rather than the Planck mass as in~\cite{Dvali:2007hz,Dvali:2007wp}.}
In the context of the Dark Dimension scenario, one can use \eqref{masstowerDD} and estimate the quantum gravity cut-off in terms of the dark energy density as
\begin{equation}\label{QGcutoff}
\Lambda_{\rm QG} = \lambda^{-1/3}\Lambda^{1/12} M_{\rm P}^{2/3} \sim 10^{9}~{\rm GeV}.
\end{equation}
In this single-tower scenario, the main formula of the Gravitino Conjecture becomes
\begin{equation}
m_1 = \lambda^{-1}_{3/2}\left(\frac{M_{3/2}}{M_{\rm P}}\right)^n M_{\rm P}\,,
\end{equation}
with $n$ still undetermined. 
Solving this expression for the gravitino mass leads to
\begin{equation}
M_{3/2}= \left(\lambda_{3/2} \ \frac{m_1}{M_{\rm P}}\right)^{\frac 1n}
M_{\rm P} = \left(\frac{\lambda_{3/2}}{\lambda}\right)^\frac{1}{n} \
\left(\frac{\Lambda}{M_{\rm P}^4}\right)^{\frac an} M_{\rm P}\,.
\label{mgravitino}
\end{equation}
We then see that, once we fix $a=1/4$ and $\lambda= 10^{-3}$, different choices of $n$ lead to different values of $M_{3/2}$. 
They are reported in Table~\ref{tabla1} at the end of this section. 
Alternatively, we can also solve the expression for the cosmological constant and, for $a=1/4$, we get
\begin{equation}
\label{Lambdacc=m32}
\Lambda = \left(\frac{\lambda}{\lambda_{3/2}}\right)^4\left(\frac{M_{3/2}}{M_{\rm P}}\right)^{4n}
\,M_{\rm P}^4\, .
 \end{equation}
This can be interpreted as the leading non-vanishing power of the supertrace ${\rm Str} {\cal M}^{4n}$, with $(\lambda/\lambda_{3/2})^4$ its corresponding leading power coefficient. 
For $n=0$, it is the partition function, which vanishes in
supersymmetric theories.
For $n=1/2$, it corresponds to the quadratic divergence of the cosmological constant.
For $n=1$, it corresponds to the logarithmic  divergence. 

More generally, since we have only one relevant mass scale, we can set up a power-series expansion for the scalar potential in terms of the small parameter $M_{3/2}/M_{\rm P}$, namely
\begin{equation}
\label{Lambdaseries}
\Lambda = M_{\rm P}^4\,\sum_{k=1}^\infty \, c_k \left(\frac{M_{3/2}}{M_{\rm P}}\right)^k. 
\end{equation}
Moreover, below we consider even integer values for $k$, implying a power series in the Newton constant $G_N\sim 1/M_{\rm P}^2$. 
Then, the relation \eqref{Lambdacc=m32} corresponds to having $k=4n$ as the leading non-vanishing term, with $n$ integer or half-integer. Indeed, the Dark Dimension fixes the ratio $a=n/k=1/4$.

In this minimal scenario, we can give bounds for the parameters $n$ and $k$. Recall that the parameter $a$ is constrained as shown in \eqref{bounda}.
Due to \eqref{nak}, this means for us that
\begin{equation}
\label{kboundn}
\frac k4 \leq n \leq \frac k2\, .
\end{equation}
As argued in \cite{Cribiori:2021gbf}, requiring that $M_{3/2}  \leq \Lambda_{\rm QG}$ results in $n \leq 3$ which, combined with the previous relation, gives $k\leq 12$. Therefore, we have 
\begin{equation}
1 \leq k \leq 12\, ,
\end{equation}
which sets the allowed interval for the power of the leading term in the series \eqref{Lambdaseries}. In the Dark Dimension, $a=n/k=1/4$ and thus we have a further restriction to
\begin{equation}
\frac14 \leq n \leq 3\, .
\end{equation}
We notice that this range for the exponent $n$ of the Gravitino Conjecture includes the lower bound $n>1/3$ found in \cite{Castellano:2021yye} via different arguments. Moreover, the upper bound was found in \cite{Cribiori:2021gbf} by assuming $\lambda_{3/2}=1$ but if we take different values we can relax the bound and go to higher values of $n$.
To summarize, we learned that swampland and unitarity bounds are powerful enough to forbid any leading term with $k>12$ in the series \eqref{Lambdaseries}.
Eventually, experimental observations will impose additional constraints, as we are going to see.

We now make a step forward and connect the discussion above to the supersymmetry breaking scale $M_{\rm SUSY}$.
One feasible possibility is that supersymmetry breaking in the bulk, related to $M_{3/2}$, is induced from a higher supersymmetry breaking scale $M_{\rm SUSY}$ on the brane, which is transmitted to the bulk via gravitational interactions that are suppressed by $M_{\rm P}$. This results into
\begin{equation}
  M_{3/2} = \varkappa \ \frac{M_{\rm SUSY}^2}{M_{\rm P}} \ , 
\label{m3229}
\end{equation}
where $\varkappa$ is an order-one parameter. 
Using such a relation, we see that, in the Dark Dimension scenario, $M_{\rm SUSY}$ is connected to the dark energy density as
\begin{equation}
M_{\rm SUSY} = \left(   \frac{\lambda_{3/2}}{\lambda\times \varkappa^n}\right)^{\frac{1}{2n}} \left(\frac{\Lambda}{M_{\rm P}^4}\right)^{\frac{1}{8n}}M_{\rm P}\, .
\end{equation}
 Once we fix $\lambda=10^{-3}$, then different choices of $n$ lead to different values of $M_{\rm SUSY}$, which are reported in Table \ref{tabla1} at the end of this section.
Alternatively, we can also write
\begin{equation}
\Lambda = \left(  \frac{\lambda\times \varkappa^n}{\lambda_{3/2}}\right)^{4}\, \left(\frac{M_{\rm SUSY}}{M_{\rm P}}\right)^{8n}\,M_{\rm P}^4\,,
\end{equation}
which corresponds to the term with exponent $k=4n$ in a more general power-series expansion in the small parameter $M_{\rm SUSY}/M_{\rm P}$, namely
\begin{equation}
\Lambda = M_{\rm P}^4\,\sum_{k=1}^\infty \, c_k' \left(\frac{M_{\rm SUSY}}{M_{\rm P}}\right)^{2k}\,,
\end{equation}
matching in fact with \eqref{Lambdaseries}, once we use \eqref{m3229}.
If we require that $M_{\rm SUSY} \leq \QG$, then we get a stronger bound on the exponent of the Gravitino Conjecture, such as $n\leq 3/2$ (this is found again assuming $\lambda_{3/2}=1$). However, in this context, we can allow for a  tuning $\varkappa^{-1}(\lambda_{3/2})^{1/2}\lesssim 10^{-4}$ such to make this scenario compatible with a maximum value $n=2$. This, using \eqref{kboundn}, truncates both the series above and \eqref{Lambdaseries} down to
\begin{equation}
1\leq k \leq 8\,.
\end{equation}
In the context of the Dark Dimension scenario ($a=n/k=1/4$), one then has
\begin{equation}
\frac14 \leq n \leq 2\,.
\end{equation}

We can now combine these theoretical predictions with experimental observations.
On the one side, $M_{\rm SUSY}$ should be lower than the quantum gravity cut-off, i.e.~$M_{\rm SUSY} < \Lambda_{\rm QG}$. In fact, we should have the hierarchy $M_{3/2}<M_{\rm SUSY}<\Lambda_{\rm QG}$.
On the other side, the scale of supersymmetry breaking is constrained from below by bounds from the Large Hadron Collider (LHC)~\cite{ParticleDataGroup:2022pth}. 
In fact, the current LHC bound on the heaviest supersymmetric particles is roughly $3~{\rm TeV}$~\cite{ParticleDataGroup:2022pth}. 
Another noteworthy point is that bounds on structure formation prevent a dark matter explanation in terms of gravitinos with $M_{3/2} < 4.7~{\rm eV}$ at 95\% CL~\cite{Osato:2016ixc}. 
Nevertheless, in the Dark Dimension scenario there are two plausible dark matter candidates: primordial black holes with Schwarzschild radius smaller than a micron~\cite{Anchordoqui:2022txe}, and massive spin-2 KK excitations of the graviton~\cite{Gonzalo:2022jac}, possibly with an interesting close relation between the two candidates~\cite{Anchordoqui:2022tgp}. 

Note that the relation of the supersymmetry breaking scale $M_{\rm SUSY}$ to the soft-terms within a supersymmetric extension of the Standard Model depends on the breaking mediation mechanism and on the particular model building arena. 
For instance, in the framework of standard {\it gravity mediation}, supersymmetry breaking occurs in a hidden sector (brane) so that soft terms are suppressed by the Planck mass, $M_{\rm soft}\sim M^2_{\rm SUSY}/M_{\rm P}$ and are thus of the order of the gravitino mass following \eqref{m3229}. In the context of the Dark Dimension, this requires a high $M_{\rm SUSY}$ of order of the string scale. On the other hand, in the framework of {\it gauge mediation}, supersymmetry breaking is mediated via gauge interactions leading to 
a gauge loop factor suppression of the soft terms, $M_{\rm soft}/M_{\rm SUSY}\sim{\cal O}(\alpha)$, with $\alpha$ an appropriate gauge coupling (assuming no other scale). Thus, $M_{\rm soft}$ is of order $(M_{3/2}M_{\rm P})^{1/2}$ and the gravitino mass is much lower.  
In general, the mediation mechanism may be more involved from these two simple cases, besides possible additional effects from the presence of the light KK tower(s), and requires a separate study that goes beyond the scope of the present paper. For instance, when the messenger mass $M_{\rm mess}$ is higher than $M_{\rm SUSY}$, the soft terms become $M_{\rm soft}\sim \alpha M^2_{\rm SUSY}/M_{\rm mess}$, connecting the two ends of gauge and gravity mediation by varying $M_{\rm mess}$.
Generically, the case $n=1$ that we propose here requires a gauge mediation type model for the soft terms, while $n=2$  might require a type of gravity mediation.

\begin{table}[ht]
  \begin{tabular}{cccc}
    \hline
    \hline
    ~~$n$~~~~ & ~~~~$k$~~~~ & ~~~~$M_{3/2}\times (\lambda_{3/2})^{-\frac 1n}$ GeV$^{-1}$~~~~ & ~~~~
    %$M_{\rm SUSY}/(\sqrt{\lambda_{3/2}^{\textcolor{red}{\frac1n}}/\varkappa}~{\rm GeV})$
    $M_{\rm SUSY} \times \varkappa^\frac12 (\lambda_{3/2})^{-\frac1{2n}}$ GeV$^{-1}$ 
    ~~\\
\hline
    $1/2$ & 2 & $~~2.5 \times 10^{-36}$    &  ~$2.5 \times 10^{-9}$   \\
    $1$ & 4 & $~2.5 \times 10^{-9}$ & $7.8 \times 10^4$ \\
    $3/2$ & 6 &  $2.5 \times 10^0$  & $2.5 \times 10^9$ \\
    $2$ & 8 & $7.8 \times 10^4$ & ~$4.4 \times 10^{11}$  \\
    \hline
    \hline
  \end{tabular}
    \caption{Gravitino mass and supersymmetry breaking scale for certain values of the parameters $n$ and $k$ in the Dark Dimension scenario, $a=1/4$. \label{tabla1}} 
  \end{table}

We conclude that the combination of experimental and
theoretical constraints leave three possibilities characterized by $n=1$, $n=3/2$ and $n=2$. The case $n=1/2$ is in fact excluded by experiment because it leads to a too low SUSY breaking scale, as reported in Table \ref{tabla1}. The case $n=1$ yields $M_{\rm SUSY}={\cal O}(\Lambda^{1/8})={\cal O}$(TeV) and, therefore, a soft mass scale of the same order (since it requires gauge mediation). More specifically, it leads to $M_{\rm SUSY} \simeq 78 \ \sqrt{\lambda_{3/2}/\varkappa}~{\rm TeV}$. Therefore, the LHC during Run III and its high-luminosity era will be able to probe part of the $\lambda_{3/2}/\varkappa$ parameter space. 
The cases $n=3/2$ and $n=2$ might be realized both in the context of gauge mediation (with an appropriate messenger mass) or in the context of gravity mediation (with an appropriate choice of the parameters $\varkappa$ and $\lambda_{3/2}$). These last two cases correspond to a high supersymmetry breaking scale $M_{\rm SUSY} \sim 10^9~{\rm GeV}$, of the order
of the string scale. The fact that the gravitino is light, ${\cal O}
({\rm GeV - TeV})$, implies that supersymmetry is still part of the effective field
theory, at least in a non-linear realization.

\subsection{Scenario with a double Kaluza-Klein tower}
\label{sec:2KKtower}

This is a more involved scenario in which there is a second tower of KK states related to ${p}$ large additional dimensions of radius ${R_2}$, in addition to the Dark Dimension with radius ${R_1 =\lambda \Lambda^{-1/4}}$.
This second KK tower has mass scale $m_2 = 1/R_2$ which, according to the Gravitino Conjecture, is related to $M_{3/2}$ as 
\begin{equation}  
M_{3/2}= \left(\lambda_{3/2} \ \frac{m_2}{M_{\rm P}}\right)^{\frac1n}M_{\rm P}
%=\left({\frac{\lambda_{3/2}}{ R_2\, M_{\rm P}}}\right)^{\frac1n} M_{\rm P}
\, ,
\label{m32212}
\end{equation}
but it is in general not related to dark energy density $\Lambda$. 
In total, there are thus $(1 + p)$ extra non-isotropic dimensions, with radii $R_1$ and $R_2$. 
The associated numbers of light species are $N_1=R_1 ~\Lambda_{\rm QG}$ and $N_2 = R_2^p ~ \Lambda_{\rm QG}^p$, giving a total number of species $N=N_1 \,N_2 = R_1 ~R_2^p  ~\Lambda_{\rm QG}^{1+p}$.  
Then, the quantum gravity cut-off follows as $ \Lambda_{\rm QG} = M_{\rm P}/ (N_1N_2)^{1/2}$, or equivalently
\begin{equation}
\Lambda_{\rm QG} = m_1^{\frac{1}{3+p}} m_2^{\frac{p}{3+p}} M_{\rm P}^{\frac{2}{3+p}},
\label{LambdaQGm1m2}
\end{equation}
which reproduces the expected behaviour for a decompactification from $4$ to $4+(1+p)$ dimensions.
We recall that $m_1 = \lambda^{-1}\,10^{-12}$ GeV and we use $\lambda \simeq 10^{-3}$, corresponding to $R_1\sim 1 \mu$m.
Furthermore, in the presence of D-branes $\Lambda_{\rm QG}$ can be related to the string scale $M_s$ via 
\begin{equation}
M_s = \alpha^\frac{2}{3+p} \Lambda_{\rm QG}\,, 
\label{Ms}
\end{equation}
where $\alpha$ is the gauge coupling of the theory on the brane worldvolume.
Notice that we are implicitly assuming that the different towers associated to $m_1$ and $m_2$ are becoming light at the same rate. This assumption can be relaxed by employing the algorithm proposed in \cite{Castellano:2021mmx} to determine the species scale.

We want now to introduce the supersymmetry breaking scale into the analysis. By recalling \eqref{m3229}, which is valid for $M_{\rm SUSY} <\Lambda_{\rm QG}$, and combining it with \eqref{m32212}, we obtain
\begin{equation}
  m_2 = \frac{\varkappa^{n}}{\lambda_{3/2}} \left(\frac{M_{\rm SUSY}}{M_{\rm P}}\right)^{2n}\, M_{\rm P} \,,
\label{M*}
\end{equation}
relating the size $R_2=1/m_2$ of the $p$ extra dimensions to $M_{\rm SUSY}$.

We look then at experimental constraints. There are essentially two inputs that we are going to use. First, data from supernova and neutron-star heating, which probe temperatures of 10~MeV, imply $1/R_2 =m_2\gtrsim 10~{\rm MeV}$~\cite{Hannestad:2003yd}. Second, LHC data that force $M_{\rm SUSY}\gtrsim 10~{\rm TeV}$. We proceed by discussing how these experimental data constrain the parameters $p$ and $n$ in several cases, setting $\varkappa=\lambda_{3/2}=\mathcal{O}(1)$ for the time being.

For $n=1/2$, the relation \eqref{M*} becomes $m_2 \simeq M_{\rm SUSY}$. In this case, the strongest bound comes from colliders and it implies also the astrophysical constraint. 
Substituting $m_2 \simeq M_{\rm SUSY} \gtrsim 10~{\rm TeV}$ into \eqref{LambdaQGm1m2}, we obtain that for $1 \leq p \leq 5$ the quantum gravity cut-off is in the range $10^7< \Lambda_{\rm QG}/{\rm GeV} \leq 10^8$. 
If we adopt the working assumption of~\cite{Antoniadis:1998ig} in which $p=2$ large extra dimensions are related to the supersymmetry breaking scale, we have then $M_s \sim 10^7~{\rm GeV}$.

For $n \geq 1$, the astrophysical bound leads to $M_{\rm SUSY}\gtrsim10^{-10/n + 18}~{\rm GeV} \gtrsim 10^8~{\rm GeV}$, which is stronger than the collider constraint. 
In this case, (for $\varkappa=\lambda_{3/2}=\mathcal{O}(1)$) $M_{\rm SUSY}$ is generically bigger than the cut-off, namely $M_{\rm SUSY}\gtrsim \Lambda_{\rm QG}$, and the relation \eqref{m3229} is not valid anymore. One should rather replace $M_{\rm SUSY}$ by $\Lambda_{\rm QG}$ therein and, after equating this new relation to \eqref{m32212}, one finds
\begin{equation}
m_2 = \frac{\varkappa^{n}}{\lambda_{3/2}} \left(\frac{\Lambda_{\rm QG}}{M_{\rm P}}\right)^{2n}\, M_{\rm P} \,.
\end{equation}
Combining $m_2$ with the expression derived from \eqref{LambdaQGm1m2}, one obtains 
\begin{equation}
 %\Lambda_{\rm QG}^3 = m_1 \ M_{\rm P}^2 \ \left(\frac{\Lambda_{\rm QG}}{M_{\rm P}} \right)^{(2n-1)p} \, .
 \Lambda_{\rm QG} = \left(\frac{\varkappa^n}{\lambda_{3/2}}\right)^{\frac{p}{3+p-2np}} \left(\frac{M_{\rm P}}{m_1}\right)^{\frac{1}{2np-(3+p)}} \,M_{\rm P}\,.
\end{equation}
Since $M_{\rm P}/m_1  =10^{27}>1$, if the parameters $n$ and $p$ are such that $2np > 3+p \geq 4$, the equation above implies $\Lambda_{\rm QG}>M_{\rm P}$ and all of models associated to these $n,p$ are excluded. 
Since $p\geq 1$, this happens for $n\geq2$. It also happens for $n=3/2$ and $p\ge 3$, as well as for $n=1$ and $p\ge 2$. 
Note that when $2np-(3+p)=0$ one obtains a contradiction $m_1=M_{\rm P}$. For the cases
$n=3/2$ and $p=1$, or $n=1$ and $p=2$, the cut-off scale becomes too low, $\Lambda_{\rm QG}\sim m_1=2.5$ eV.
Still, a possibility seems to be $n=1=p$, for which $\Lambda_{\rm QG} = \sqrt{\varkappa/\lambda_{3/2}} \times 78$ TeV and $M_{3/2} =\lambda_{3/2} m_2 = \varkappa \times 2.5$ eV, but this latter value is not compatible with the astrophysical bound. 

Thus, all values $n\geq 1$ seems to be excluded if $\varkappa=\lambda_{3/2}=\mathcal{O}(1)$. However, in the case $n=1=p$, one can try to tune the ratio $\lambda_{3/2}/\varkappa =\mathcal{O}(10^{-5})$, to obtain the hierarchy $M_{\rm SUSY} \lesssim \Lambda_{\rm QG}$ and remain in the regime of validity of \eqref{m3229}.  For $p>1$ the value of the
cut-off becomes smaller, requiring a smaller value of $\lambda_{3/2}/\varkappa$.

In summary, we conclude that the combination of experimental and
theoretical constraints leave two possibilities. The first is characterized by $n=1/2$ and $1\leq p \leq 5$. The supersymmetry breaking scale can be as low as the present experimental bounds, $M_{\rm SUSY} \sim m_2 > 10~{\rm TeV}$, implying that the gravitino mass $M_{3/2} > 0.1~{\rm eV}$, while the quantum gravity cut-off (string scale) varies between $10^7$ to $10^8~{\rm GeV}$. 
The second possibility is for $n= 1=p$ and the supersymmetry breaking scale is high, near the string scale $M_s\sim \Lambda_{\rm QG}\sim 10^7~{\rm GeV}$. 
One can tune $\lambda_{3/2}/\varkappa =\mathcal{O}(10^{-5})$ in order to accommodate $m_2 \sim M_{3/2}/\lambda_{3/2} \sim 10~{\rm MeV}$. 

\section{String and effective models}
\label{sec:eftmodels}

In this section, we investigate how the Dark Dimension scenario together with supersymmetry breaking can be realized in effective supergravity descriptions of string compactifications. 
Concretely, we concentrate on the scenario with a single KK tower of states presented in section \ref{sec:1KKtower}.
The other scenario with two KK towers, presented in section \ref{sec:2KKtower}, is more involved.

The effective models of this section are to be intended as a proof of principle that these scenarios can find an embedding into string theory. In particular, they reproduce the correct parametric relations among the different energy scales, such as dark energy density, supersymmetry breaking scale and KK scale. More importantly, they lead to a decompactification limit with $\Lambda\sim R_1^{-4}$. This represents an upper bound for the potential even in the case one turns out to be able to construct more realistic models properly realising a cosmological constant or a quintessence phase. However, in the present work, we do not address the problem of finding an appropriate radius stabilization mechanism, for it is not addressed in the original Dark Dimension scenario \cite{Montero:2022prj} either. Therefore, in what follows, we just comment on the possible string realization of the discussed scenarios, while we leave the explicit construction of an effective supergravity or superstring model with possibly all moduli stabilized for future work.

\subsection{String realizations}

From the analysis in section~\ref{sec:GCC}, we obtained five viable solutions, corresponding to $n=1$ for both the scenarios with a single and a double KK-towers, $n=1/2$ for the scenario with a double KK-tower and $n=3/2$ or $n=2$ again for the scenario with a single tower, namely that of section \ref{sec:1KKtower}.
In the first three of these solutions, the gravitino mass or the supersymmetry breaking scale is proportional to a compactification scale. 
In the remaining two solutions, the supersymmetry breaking scale
is instead higher and of order the string scale, here intended as the quantum gravity cutoff.

The first three models can be realized in string theory by imposing Scherk-Schwarz boundary conditions where the fields of the higher dimensional theory are taken to periodic up to an ${\cal R}$-symmetry transformation, which breaks supersymmetry \cite{Scherk:1978ta,Scherk:1979zr}. 
This amounts to a corresponding shift of the KK momentum. 
In particular, for the models of sections~\ref{sec:1KKtower} and~\ref{sec:2KKtower} with $n=1$, the gravitino zero-mode acquires a mass~\cite{Rohm:1983aq,Kounnas:1988ye,Ferrara:1988jx,Antoniadis:1990ew} 
\begin{equation}
  M_{3/2} = \frac{q}{R} \,,
\label{q}
\end{equation}  
where $q$ is proportional to the ${\cal R}$ charge, which is a quantized order-one parameter. 
Here, $R$ can be either of the radii $R_1$ or $R_2$ introduced in section~\ref{sec:Scenarios}. 
In these examples, we can identify the parameter entering the Gravitino Conjecture as $\lambda_{3/2} = q$. 
The simplest string realization consists of temperature-like boundary conditions, where fermions are anti-periodic and their KK momenta become half-integers, yielding $q = 1/2$. 
The effective supergravity has been worked out in \cite{Antoniadis:1991kh} and the construction has been extended to models with
branes in \cite{Antoniadis:1998ki}. 
It turns out that when these boundary conditions are imposed in a direction which is transverse to the brane, there is no tree level effect on the brane.
On the other hand, if the boundary conditions are imposed in a direction along the brane, the bulk is then affected.
It remains to be found a way to implement the breaking on the brane consistently with the breaking in the bulk, in order to match with \eqref{m3229}.   
For the model of section~\ref{sec:2KKtower} with $n=1/2$ and $1\leq p \leq 5$, a relation similar to \eqref{q}  holds for the scale of supersymmetry breaking, with $M_{\rm SUSY}$ characterizing the mass of gauginos and/or sfermions.

Finally, if the supersymmetry breaking scale is of the order of the string scale, as for $n=3/2$ or $n=2$ of section~\ref{sec:1KKtower}, the relation \eqref{m3229} is realized in the framework of brane supersymmetry breaking, where the bulk remains supersymmetric to lowest order and the breaking is transmitted via gravitational interactions~\cite{Antoniadis:1999xk}. 
Then, as a consequence of swampland conjectures, in these models one expects a tower of states with $n=3/2$ or $n=2$.

\subsection{Low energy effective models}

We consider spontaneous F-term supersymmetry breaking in the context of four-dimensional ${\cal N}=1$ supergravity, specified by a real K\"ahler potential $K(\phi,\bar\phi)$ and an holomorphic superpotential $W(\phi)$, with $\phi$ being chiral multiplets.
The scalar potential $V\equiv \Lambda$ and the gravitino mass $M_{3/2}$ then take their well-known schematic forms
\begin{eqnarray}\label{scalarpotential}
V= M_{\rm P}^2~e^K(|DW|^2-3|W|^2)\, ,\qquad M_{3/2}^2=e^K|W|^2\, .
\end{eqnarray}
In our conventions, the superpotential has mass dimension 1, while the K\"ahler potential and the scalar fields $\phi$ are dimensionless; in particular, the latter are given in units of the string scale $M_s$ (or of the ten-dimensional Planck mass, $M_{\rm P}^{(10)}$).

As mentioned in section \ref{sec:GCC}, for supersymmetric anti-de Sitter minima, $DW=0$ and one gets the relation \eqref{m32SUSYAdS}, satisfying $n=2a$, i.e.~$k=2$.
For non-supersymmetric backgrounds with positive energy density $\Lambda$, one can envisage a general power-series relation as in \eqref{Lambdaseries}. (We are here considering only the scenario with a single KK-tower.)
As discussed in section \eqref{sec:1KKtower}, the lowest possible term with $k=2$ is ruled out by experiments, since it leads to a supersymmetry breaking scale which is too low. 
Therefore, we look for effective theories with $c_2=0$ and first non-vanishing coefficient $c_4\neq 0$. 
These can be realized in four-dimensional ${\cal N}=1$ effective
models arising from compactifying heterotic or type II string on Calabi-Yau (orientifolds) or \`a la
Scherk-Schwarz~\cite{Scherk:1978ta,Scherk:1979zr} with appropriate boundary conditions~\cite{Rohm:1983aq,Kounnas:1988ye,Ferrara:1988jx,Antoniadis:1990ew}.

To describe the class of low energy models, we introduce three chiral moduli fields $\phi_i$, with $i=1,2,3$. 
The superpotential $W\simeq {\cal O}(M_{\rm P})$ is assumed to be constant and we have a no-scale structure at tree level, which is broken by a one loop (or ${\alpha'}^3$) correction $\xi$ to the K\"ahler potential
\begin{equation}
K =-\log \left((-i(\phi-\bar \phi))^3+\xi\right)\, .
\end{equation}
Here and in the following, to avoid cluttering various formulae, we employ the shorthand notation $\prod_{i=1}^3(-i(\phi_i-\bar \phi_i))\equiv (-i(\phi-\bar \phi))^3 \equiv (2\,{\rm Im}\phi)^3$ and we generically set the axions to zero.
Then, the gravitino mass is
\begin{equation}
M_{3/2} = \frac{|W|}{\sqrt{ (2{\rm Im}\, \phi)^3+\xi}} = \frac{|W|}{(2{\rm Im}\, \phi)^\frac32} + \mathcal{O}(\xi)\label{m32}
\end{equation}
and the scalar potential becomes
\begin{equation}
V=6M_{\rm P}^2\xi \frac{ |W |^2}{(2{\rm Im} \phi)^6} + \mathcal{O}(\xi^2) = \frac{6M_{\rm P}^2\xi}{|W |^2 }( M_{3/2})^4 + \mathcal{O}(\xi^2).\label{Lambda}
\end{equation}
By comparing these with equations \eqref{mgravitino} and \eqref{Lambdaseries}, we see that for this class of effective models we indeed get $n=4a$, i.e.~$k=4$ and $c_4=\left(\frac{\lambda}{\lambda_{3/2}}\right)^4=\frac{6M_{\rm P}^2\xi}{|W |^2}\neq 0$.

Now we study for which values of $a$ and $n$ these models can be accommodated into a compactification of string theory down to four dimensions and with moduli $\phi$.
Since we want to realize the Dark Dimension scenario, we have to assume that the compact space is non-isotropic with one large dimension of size $R_1$ and the other five dimensions of string size. 
To match with the language of string compactifications, in which moduli are typically dimensionless, we introduce dimensionless radii $r_i$, associated to the dimensionful quantities $R_i$.
In general, we expect that  ${\rm Im }\phi$ scales with $r_1$ as 
\begin{equation}
{\rm Im} \phi\sim (r_1)^{\beta/3}\, ,\label{alpha}
\end{equation}
where the parameter $\beta$ is to be determined from the specific compactification scheme under investigation.

We consider three types of compactifications leading to four-dimensional $\mathcal{N}=1$ supergravity at low energies: $(i)$ heterotic or type IIA orientifolds on a Calabi-Yau, $(ii)$ type IIB on a Calabi-Yau orientifold and $(iii)$ heterotic or type II compactified \`a la Scherk-Schwarz. We discuss these more in detail below.

\begin{itemize}
\item[$(i)$] \emph{Heterotic or type IIA orientifolds on Calabi-Yau}.\\
In this case, the moduli $\phi_i$ correspond to volumes of 2-cyles of the Calabi-Yau three-fold, while $({\rm Im}\phi)^3\simeq {\cal V}$ is the full volume. 
We assume that all other moduli as well as the four-dimensional dilaton are already stabilized or projected out.\footnote{We recall that the four-dimensional dilaton is given as $e^{-2\phi_{(4)}} = \mathcal{V}\, e^{-2\phi_{(10)}}$.} 
Since there is one large direction, we get that ${\cal V}\simeq r_1$ and hence ${\rm Im}\phi\simeq (r_1)^{1/3}$, i.e.~$\beta=1$, leading to
\begin{equation}
M_{3/2} \simeq {\frac{M_{\rm P}}{(r_1)^{1/2}}} \, ,\qquad
V\simeq {\frac{M_{\rm P}^4}{(r_1)^2}}\, .
\end{equation}

\item[$(ii)$] \emph{Type IIB on Calabi-Yau orientifolds with three K\"ahler moduli $\phi_i$}.\\
In this case, the fields $\phi_i$ correspond to volumes of 4-cycles of the Calabi-Yau, while $({\rm Im}\phi)^3\simeq {\cal V}^2$ is the square of the full volume. 
We assume that all other moduli as well as the four-dimensional dilaton are already stabilized or projected out. Since there is one large direction, we get that
${\cal V}\simeq r_1$ and hence ${\rm Im}\phi\simeq (r_1)^{2/3}$, i.e. $\beta=2$,
 leading to
\begin{equation}
M_{3/2} \simeq {\frac{M_{\rm P}}{ r_1}} \, ,\qquad
V\simeq {\frac{M_{\rm P}^4}{ (r_1)^4}}\, .
\end{equation}

\item[$(iii)$] \emph{Heterotic and type II STU models with three fields $\phi_i$ (Scherk-Schwarz)}.\\
In this case, the field $\phi_1\equiv T$ is the K\"ahler modulus $T$ of a two-torus, the field $\phi_2\equiv U$ is the corresponding complex structure modulus, and the field $\phi_3\equiv S$ is the four-dimensional heterotic dilaton.
Then, we get that ${\rm Im}T\simeq r_1$, ${\rm Im}U\simeq r_1$, and ${\rm Im}S\simeq r_1$. Since there is one large direction, we get that $\mathcal{V}\simeq r_1$ and hence $({\rm Im}\phi)^3 ={\rm Im}S\, {\rm Im}T\, {\rm Im}U\simeq r_1^3$, i.e.~$\beta=3$, leading to
\begin{equation}
M_{3/2} \simeq {\frac{M_{\rm P}}{ (r_1)^{3/2}}} \, ,\qquad
V\simeq {\frac{M_{\rm P}^4}{ (r_1)^6}}\, .
\end{equation}

\end{itemize}

We can now compare the behaviour of $M_{3/2}$ and $V\equiv \Lambda$ in these models with the mass scale of the relevant tower of KK states. As we will see, only one out of the three string realizations discussed above can accommodate the Dark Dimension scenario.

In general, the lowest KK mode corresponding to one direction of the compact six-dimensional space with radius $r_1$ is given in string units as
\begin{equation}
m_{KK}= {\frac{M_s}{r_1}}\, .
\end{equation}
Recall however that in the effective supergravity description all masses are measured in units of the four-dimensional Planck mass $M_{\rm P}$.
Thus, when passing from string to Planck units,  $M_s \simeq M_{\rm P}/\sqrt{\cal V}$, we get for one large dimension an additional factor $1/\sqrt{\mathcal{V}}=1/\sqrt{r_1}$, such that the KK scale becomes
\begin{equation}
m_{KK}= {\frac{M_{\rm P}}{ (r_1)^{3/2}}} \, .\label{mkk}
\end{equation} 
Comparing the KK mass in \eqref{mkk} with $M_{3/2}$ in \eqref{m32} and with the scaling of the moduli \eqref{alpha}, we deduce that the parameter $n$ of the Gravitino Conjecture is given by
\begin{equation}
n={\frac{3}{ \beta}}\, .
\end{equation}
Similarly, comparing $V\equiv \Lambda$ in \eqref{Lambda} with the KK scale \eqref{mkk}, we derive that the parameter $a$ of the ADC is given as
\begin{equation}
a={\frac{3 }{4\beta}}\, .
\end{equation}
As a result, we see that we need $\beta=3$ in order to realize the Dark Dimension with $a=1/4$. This means that the compactification scheme $(iii)$, namely the $STU$ model,
provides a concrete string realization of the Dark Dimension scenario. This corresponds to the effective field theory of the Scherk-Schwarz string realization~\cite{Antoniadis:1991kh}.

Finally, recall that swampland and unitary constraints restrict the parameter $a$ as in \eqref{bounda}. This translates into a bound for $\beta$ of the type
\begin{equation}
\frac32\leq \beta\leq 3\, ,
\end{equation}
which is satisfied by the compactification schemes $(ii)$ and $(iii)$, but not by $(i)$.

\section{Conclusions}
\label{sec:concl}

In this paper, we derived a relation between the dark energy density $\Lambda$ and the supersymmetry breaking scale $M_{\rm SUSY}$, in the framework of the recently proposed Dark Dimension scenario \cite{Montero:2022prj}. Employing the value of $\Lambda$ measured today, we are led to a prediction of the supersymmetry breaking scale of the order $M_{\rm SUSY}={\cal O}(\Lambda^{1/8})=\mathcal{O}(10-100)$ TeV, in the most favorable case. Depending on the choice of few parameters, a higher supersymmetry breaking scale is possible as well, leading instead to a gravitino mass \mbox{$M_{3/2}=\mathcal{O}({\rm GeV - TeV})$}. Both these energy scales are within reach of current and next generation of colliders.

We arrived at this result by combining two swampland conjectures, such as the Anti-de Sitter Distance Conjecture on $\Lambda$ \cite{Lust:2019zwm} and the Gravitino Conjecture on the gravitino mass $M_{3/2}$ \cite{Cribiori:2021gbf,Castellano:2021yye}. Indeed, in a quasi-flat background, the gravitino mass becomes simply a measure of the supersymmetry breaking scale . These conjectures state that in the limit of small $\Lambda$ or $M_{3/2}$, respectively, a tower of Kaluza-Klein states becomes light.

Hence, we discussed two main scenarios. In the first, the dark energy scale $\Lambda$ and the gravitino mass $M_{3/2}$ are related to the same tower of Kaluza-Klein states, fixing also the size of the Dark Dimension. In the second, the towers are different, but $\Lambda$ is again related to the Dark Dimension. The first scenario is arguably the simplest and it is the one from which we extracted our main predictions. The phenomenological implications of the second scenario have been discussed to some extent as well, but we left the construction of explicit models for future work.

In the minimal scenario with a single Kaluza-Klein tower of states, we related to one another the parameters $a$ and $n$ determining, respectively, the scaling of $\Lambda$ and $M_{3/2}$ with respect to the Kaluza-Klein mass. 
Eventually, their ratio has to be a model-dependent constant, $k=n/a$, which enters a general power-series expansion of $\Lambda$ in terms of $M_{3/2}$, or equivalently of $M_{\rm SUSY}$. In turn, this can be rephrased in terms of the supertrace series $\Lambda=\sum_k{\rm Str} {\cal M}^{k}$.

The free parameters can be fixed by enforcing $a=1/4$, as required by the Dark Dimension, and by looking at experimental bounds. 
As a consequence, one finds that the case $n=1/2$ (corresponding to $k=2$) is excluded, as it leads to a too low scale of supersymmetry breaking. 
However, the case $n=1$ (corresponding to $k=4$) is the first phenomenologically viable option and it leads to the intriguing swampland prediction that $M_{\rm SUSY}={\cal O}(\Lambda^{1/8})={\cal O}({\rm TeV})$. 
This prediction and (part of) the associated parameter space for the parameters  $\lambda_{3/2}$ and $\varkappa$ can then be tested by the future runs of LHC or in the next generation of high energy particle colliders.
For this case, we also identified a possible string embedding in terms of Scherk-Schwarz compactifications~\cite{Antoniadis:1991kh}.
Higher values of $n$ are in principle also possible, such as $n=3/2$ or $n=2$. They lead to a supersymmetry breaking scale near the string scale and gravitino mass $M_{3/2}=\mathcal{O}({\rm GeV - TeV})$.
Therefore, all of these scenarios could potentially rekindle the search for near-by supersymmetry at the LHC.

As already said, we identified certain Scherk-Schwarz string compactifications as possible candidates to provide 
realizations of the Dark Dimension scenario ($a=1/4$) in the case $n=1$ (i.e., $k=4$). Another candidate with $a=1/4$ has been recently identified by \cite{Blumenhagen:2022zzw} as the geometry associated to a strongly warped throat in the framework of the so-called KKLT scenario.
Here, one starts from a supersymmetric anti-de Sitter vacuum with $a=1/4$ and $k=2$. It is then assumed that the vacuum energy $|\Lambda_{\rm AdS}|=3M_{3/2}^2 M_{\rm P}^2$, the uplift potential $V_{\rm up}$, and the de Sitter cosmological constant $\Lambda=\Lambda_{\rm AdS}+V_{\rm up}$ are all of the same order. Hence, one generically expects that also after the uplift $\Lambda\sim M_{3/2}^2 M_{\rm P}^2$. This corresponds to $k=2$ and, as such, within our framework, it seems to be experimentally ruled out.

Let us conclude by mentioning that, in the context of the Swampland program, the dark energy density $\Lambda$ has been related also to other low energy observables. This is the case for the neutrino mass, $m_\nu={\cal O}(\Lambda^{1/4})$ \cite{Ibanez:2017kvh}, for the dark matter mass in the Dark Universe, $m_{DM}={\cal O}(\Lambda^{1/6})$ \cite{Gonzalo:2022jac}, and for primordial black holes as dark matter candidates with Schwarzschild radii  $r_s<{\cal O}(\Lambda^{-1/4})$ and Hawking temperature $T_H>{\cal O}(\Lambda^{1/4})$ \cite{Anchordoqui:2022txe,Anchordoqui:2022tgp}.
The implications of the remarkable way in which swampland criteria relate quantum gravity to the observable world definitely  deserve further investigations.

\vspace{0.4cm}
\noindent{\bf Acknowledgments.} 
We thank O.~Aharony, R.~Blumenhagen, M.~Cicoli, L.~Randall and C. Vafa for useful discussions.
The work of L.A.A. is supported by the U.S.~National Science Foundation (NSF Grant PHY-2112527).
The work of N.C.~is supported by the Alexander-von-Humboldt foundation.
The work of D.L.~is supported by the Origins Excellence Cluster and by the German-Israel-Project (DIP) on Holography and the Swampland.

\bibliographystyle{JHEP}
\bibliography{refGDD.bib}

\providecommand{\href}[2]{#2}\begingroup\raggedright\begin{thebibliography}{10}

\bibitem{Planck:2018vyg}
{\scshape Planck} collaboration, N.~Aghanim et~al., \emph{{Planck 2018 results.
  VI. Cosmological parameters}},
  \href{http://dx.doi.org/10.1051/0004-6361/201833910}{\emph{Astron.
  Astrophys.} {\bf 641} (2020) A6},
  [\href{https://arxiv.org/abs/1807.06209}{{\tt 1807.06209}}].

\bibitem{SupernovaCosmologyProject:1998vns}
{\scshape Supernova Cosmology Project} collaboration, S.~Perlmutter et~al.,
  \emph{{Measurements of $\Omega$ and $\Lambda$ from 42 high redshift
  supernovae}}, \href{http://dx.doi.org/10.1086/307221}{\emph{Astrophys. J.}
  {\bf 517} (1999) 565--586},
  [\href{https://arxiv.org/abs/astro-ph/9812133}{{\tt astro-ph/9812133}}].

\bibitem{SupernovaSearchTeam:1998fmf}
{\scshape Supernova Search Team} collaboration, A.~G. Riess et~al.,
  \emph{{Observational evidence from supernovae for an accelerating universe
  and a cosmological constant}},
  \href{http://dx.doi.org/10.1086/300499}{\emph{Astron. J.} {\bf 116} (1998)
  1009--1038}, [\href{https://arxiv.org/abs/astro-ph/9805201}{{\tt
  astro-ph/9805201}}].

\bibitem{Montero:2022prj}
M.~Montero, C.~Vafa and I.~Valenzuela, \emph{{The Dark Dimension and the
  Swampland}},  \href{https://arxiv.org/abs/2205.12293}{{\tt 2205.12293}}.

\bibitem{Vafa:2005ui}
C.~Vafa, \emph{{The String landscape and the swampland}},
  \href{https://arxiv.org/abs/hep-th/0509212}{{\tt hep-th/0509212}}.

\bibitem{Palti:2019pca}
E.~Palti, \emph{{The Swampland: Introduction and Review}},
  \href{http://dx.doi.org/10.1002/prop.201900037}{\emph{Fortsch. Phys.} {\bf
  67} (2019) 1900037}, [\href{https://arxiv.org/abs/1903.06239}{{\tt
  1903.06239}}].

\bibitem{vanBeest:2021lhn}
M.~van Beest, J.~Calder\'on-Infante, D.~Mirfendereski and I.~Valenzuela,
  \emph{{Lectures on the Swampland Program in String Compactifications}},
  \href{https://arxiv.org/abs/2102.01111}{{\tt 2102.01111}}.

\bibitem{Agmon:2022thq}
N.~B. Agmon, A.~Bedroya, M.~J. Kang and C.~Vafa, \emph{{Lectures on the string
  landscape and the Swampland}},  \href{https://arxiv.org/abs/2212.06187}{{\tt
  2212.06187}}.

\bibitem{Lust:2019zwm}
D.~L{\"u}st, E.~Palti and C.~Vafa, \emph{{AdS and the Swampland}},
  \href{http://dx.doi.org/10.1016/j.physletb.2019.134867}{\emph{Phys. Lett. B}
  {\bf 797} (2019) 134867}, [\href{https://arxiv.org/abs/1906.05225}{{\tt
  1906.05225}}].

\bibitem{Anchordoqui:2022ejw}
L.~A. Anchordoqui, \emph{{Dark dimension, the swampland, and the origin of
  cosmic rays beyond the Greisen-Zatsepin-Kuzmin barrier}},
  \href{http://dx.doi.org/10.1103/PhysRevD.106.116022}{\emph{Phys. Rev. D} {\bf
  106} (2022) 116022}, [\href{https://arxiv.org/abs/2205.13931}{{\tt
  2205.13931}}].

\bibitem{Anchordoqui:2022txe}
L.~A. Anchordoqui, I.~Antoniadis and D.~L{\"u}st, \emph{{Dark dimension, the
  swampland, and the dark matter fraction composed of primordial black holes}},
  \href{http://dx.doi.org/10.1103/PhysRevD.106.086001}{\emph{Phys. Rev. D} {\bf
  106} (2022) 086001}, [\href{https://arxiv.org/abs/2206.07071}{{\tt
  2206.07071}}].

\bibitem{Blumenhagen:2022zzw}
R.~Blumenhagen, M.~Brinkmann and A.~Makridou, \emph{{The Dark Dimension in a
  Warped Throat}},  \href{https://arxiv.org/abs/2208.01057}{{\tt 2208.01057}}.

\bibitem{Gonzalo:2022jac}
E.~Gonzalo, M.~Montero, G.~Obied and C.~Vafa, \emph{{Dark Dimension Gravitons
  as Dark Matter}},  \href{https://arxiv.org/abs/2209.09249}{{\tt 2209.09249}}.

\bibitem{Anchordoqui:2022tgp}
L.~Anchordoqui, I.~Antoniadis and D.~L{\"u}st, \emph{{The Dark Universe:
  Primordial Black Hole $\leftrightharpoons$ Dark Graviton Gas Connection}},
  \href{https://arxiv.org/abs/2210.02475}{{\tt 2210.02475}}.

\bibitem{Anchordoqui:2022svl}
L.~A. Anchordoqui, I.~Antoniadis and D.~L{\"u}st, \emph{{Aspects of the Dark
  Dimension in Cosmology}},  \href{https://arxiv.org/abs/2212.08527}{{\tt
  2212.08527}}.

\bibitem{Banks:2000fe}
T.~Banks, \emph{{Cosmological breaking of supersymmetry?}},
  \href{http://dx.doi.org/10.1142/S0217751X01003998}{\emph{Int. J. Mod. Phys.
  A} {\bf 16} (2001) 910--921},
  [\href{https://arxiv.org/abs/hep-th/0007146}{{\tt hep-th/0007146}}].

\bibitem{ParticleDataGroup:2022pth}
{\scshape Particle Data Group} collaboration, R.~L. Workman et~al.,
  \emph{{Review of Particle Physics}},
  \href{http://dx.doi.org/10.1093/ptep/ptac097}{\emph{PTEP} {\bf 2022} (2022)
  083C01}.

\bibitem{Antoniadis:1988jn}
I.~Antoniadis, C.~Bachas, D.~C. Lewellen and T.~N. Tomaras, \emph{{On
  Supersymmetry Breaking in Superstrings}},
  \href{http://dx.doi.org/10.1016/0370-2693(88)90679-X}{\emph{Phys. Lett. B}
  {\bf 207} (1988) 441--446}.

\bibitem{Cribiori:2021gbf}
N.~Cribiori, D.~L{\"u}st and M.~Scalisi, \emph{{The gravitino and the
  swampland}}, \href{http://dx.doi.org/10.1007/JHEP06(2021)071}{\emph{JHEP}
  {\bf 06} (2021) 071}, [\href{https://arxiv.org/abs/2104.08288}{{\tt
  2104.08288}}].

\bibitem{Castellano:2021yye}
A.~Castellano, A.~Font, A.~Herraez and L.~E. Ib\'a\~nez, \emph{{A gravitino
  distance conjecture}},
  \href{http://dx.doi.org/10.1007/JHEP08(2021)092}{\emph{JHEP} {\bf 08} (2021)
  092}, [\href{https://arxiv.org/abs/2104.10181}{{\tt 2104.10181}}].

\bibitem{Cribiori:2020use}
N.~Cribiori, G.~Dall'agata and F.~Farakos, \emph{{Weak gravity versus de
  Sitter}}, \href{http://dx.doi.org/10.1007/JHEP04(2021)046}{\emph{JHEP} {\bf
  04} (2021) 046}, [\href{https://arxiv.org/abs/2011.06597}{{\tt 2011.06597}}].

\bibitem{DallAgata:2021nnr}
G.~Dall'Agata, M.~Emelin, F.~Farakos and M.~Morittu, \emph{{The unbearable
  lightness of charged gravitini}},
  \href{http://dx.doi.org/10.1007/JHEP10(2021)076}{\emph{JHEP} {\bf 10} (2021)
  076}, [\href{https://arxiv.org/abs/2108.04254}{{\tt 2108.04254}}].

\bibitem{Palti:2020tsy}
E.~Palti, \emph{{Fermions and the Swampland}},
  \href{http://dx.doi.org/10.1016/j.physletb.2020.135617}{\emph{Phys. Lett. B}
  {\bf 808} (2020) 135617}, [\href{https://arxiv.org/abs/2005.08538}{{\tt
  2005.08538}}].

\bibitem{Coudarchet:2021qwc}
T.~Coudarchet, E.~Dudas and H.~Partouche, \emph{{Geometry of orientifold vacua
  and supersymmetry breaking}},
  \href{http://dx.doi.org/10.1007/JHEP07(2021)104}{\emph{JHEP} {\bf 07} (2021)
  104}, [\href{https://arxiv.org/abs/2105.06913}{{\tt 2105.06913}}].

\bibitem{Lee:2018urn}
S.-J. Lee, W.~Lerche and T.~Weigand, \emph{{Tensionless Strings and the Weak
  Gravity Conjecture}},
  \href{http://dx.doi.org/10.1007/JHEP10(2018)164}{\emph{JHEP} {\bf 10} (2018)
  164}, [\href{https://arxiv.org/abs/1808.05958}{{\tt 1808.05958}}].

\bibitem{Lee:2019wij}
S.-J. Lee, W.~Lerche and T.~Weigand, \emph{{Emergent strings from infinite
  distance limits}},
  \href{http://dx.doi.org/10.1007/JHEP02(2022)190}{\emph{JHEP} {\bf 02} (2022)
  190}, [\href{https://arxiv.org/abs/1910.01135}{{\tt 1910.01135}}].

\bibitem{Dvali:2007hz}
G.~Dvali, \emph{{Black Holes and Large N Species Solution to the Hierarchy
  Problem}}, \href{http://dx.doi.org/10.1002/prop.201000009}{\emph{Fortsch.
  Phys.} {\bf 58} (2010) 528--536},
  [\href{https://arxiv.org/abs/0706.2050}{{\tt 0706.2050}}].

\bibitem{Dvali:2007wp}
G.~Dvali and M.~Redi, \emph{{Black Hole Bound on the Number of Species and
  Quantum Gravity at LHC}},
  \href{http://dx.doi.org/10.1103/PhysRevD.77.045027}{\emph{Phys. Rev. D} {\bf
  77} (2008) 045027}, [\href{https://arxiv.org/abs/0710.4344}{{\tt
  0710.4344}}].

\bibitem{Hannestad:2003yd}
S.~Hannestad and G.~G. Raffelt, \emph{{Supernova and neutron star limits on
  large extra dimensions reexamined}},
  \href{http://dx.doi.org/10.1103/PhysRevD.69.029901}{\emph{Phys. Rev. D} {\bf
  67} (2003) 125008}, [\href{https://arxiv.org/abs/hep-ph/0304029}{{\tt
  hep-ph/0304029}}].

\bibitem{Kapner:2006si}
D.~J. Kapner, T.~S. Cook, E.~G. Adelberger, J.~H. Gundlach, B.~R. Heckel, C.~D.
  Hoyle et~al., \emph{{Tests of the gravitational inverse-square law below the
  dark-energy length scale}},
  \href{http://dx.doi.org/10.1103/PhysRevLett.98.021101}{\emph{Phys. Rev.
  Lett.} {\bf 98} (2007) 021101},
  [\href{https://arxiv.org/abs/hep-ph/0611184}{{\tt hep-ph/0611184}}].

\bibitem{Lee:2020zjt}
J.~G. Lee, E.~G. Adelberger, T.~S. Cook, S.~M. Fleischer and B.~R. Heckel,
  \emph{{New Test of the Gravitational $1/r^2$ Law at Separations down to 52
  $\mu$m}}, \href{http://dx.doi.org/10.1103/PhysRevLett.124.101101}{\emph{Phys.
  Rev. Lett.} {\bf 124} (2020) 101101},
  [\href{https://arxiv.org/abs/2002.11761}{{\tt 2002.11761}}].

\bibitem{Hebecker:2018vxz}
A.~Hebecker and T.~Wrase, \emph{{The Asymptotic dS Swampland Conjecture - a
  Simplified Derivation and a Potential Loophole}},
  \href{http://dx.doi.org/10.1002/prop.201800097}{\emph{Fortsch. Phys.} {\bf
  67} (2019) 1800097}, [\href{https://arxiv.org/abs/1810.08182}{{\tt
  1810.08182}}].

\bibitem{Blumenhagen:2019vgj}
R.~Blumenhagen, M.~Brinkmann and A.~Makridou, \emph{{Quantum Log-Corrections to
  Swampland Conjectures}},
  \href{http://dx.doi.org/10.1007/JHEP02(2020)064}{\emph{JHEP} {\bf 02} (2020)
  064}, [\href{https://arxiv.org/abs/1910.10185}{{\tt 1910.10185}}].

\bibitem{Castellano:2022bvr}
A.~Castellano, A.~Herr\'aez and L.~E. Ib\'a\~nez, \emph{{The Emergence Proposal
  in Quantum Gravity and the Species Scale}},
  \href{https://arxiv.org/abs/2212.03908}{{\tt 2212.03908}}.

\bibitem{Osato:2016ixc}
K.~Osato, T.~Sekiguchi, M.~Shirasaki, A.~Kamada and N.~Yoshida,
  \emph{{Cosmological Constraint on the Light Gravitino Mass from CMB Lensing
  and Cosmic Shear}},
  \href{http://dx.doi.org/10.1088/1475-7516/2016/06/004}{\emph{JCAP} {\bf 06}
  (2016) 004}, [\href{https://arxiv.org/abs/1601.07386}{{\tt 1601.07386}}].

\bibitem{Castellano:2021mmx}
A.~Castellano, A.~Herr\'aez and L.~E. Ib\'a\~nez, \emph{{IR/UV Mixing, Towers
  of Species and Swampland Conjectures}},
  \href{https://arxiv.org/abs/2112.10796}{{\tt 2112.10796}}.

\bibitem{Antoniadis:1998ig}
I.~Antoniadis, N.~Arkani-Hamed, S.~Dimopoulos and G.~R. Dvali, \emph{{New
  dimensions at a millimeter to a Fermi and superstrings at a TeV}},
  \href{http://dx.doi.org/10.1016/S0370-2693(98)00860-0}{\emph{Phys. Lett. B}
  {\bf 436} (1998) 257--263}, [\href{https://arxiv.org/abs/hep-ph/9804398}{{\tt
  hep-ph/9804398}}].

\bibitem{Scherk:1978ta}
J.~Scherk and J.~H. Schwarz, \emph{{Spontaneous Breaking of Supersymmetry
  Through Dimensional Reduction}},
  \href{http://dx.doi.org/10.1016/0370-2693(79)90425-8}{\emph{Phys. Lett. B}
  {\bf 82} (1979) 60--64}.

\bibitem{Scherk:1979zr}
J.~Scherk and J.~H. Schwarz, \emph{{How to Get Masses from Extra Dimensions}},
  \href{http://dx.doi.org/10.1016/0550-3213(79)90592-3}{\emph{Nucl. Phys. B}
  {\bf 153} (1979) 61--88}.

\bibitem{Rohm:1983aq}
R.~Rohm, \emph{{Spontaneous Supersymmetry Breaking in Supersymmetric String
  Theories}}, \href{http://dx.doi.org/10.1016/0550-3213(84)90007-5}{\emph{Nucl.
  Phys. B} {\bf 237} (1984) 553--572}.

\bibitem{Kounnas:1988ye}
C.~Kounnas and M.~Porrati, \emph{{Spontaneous Supersymmetry Breaking in String
  Theory}}, \href{http://dx.doi.org/10.1016/0550-3213(88)90153-8}{\emph{Nucl.
  Phys. B} {\bf 310} (1988) 355--370}.

\bibitem{Ferrara:1988jx}
S.~Ferrara, C.~Kounnas, M.~Porrati and F.~Zwirner, \emph{{Superstrings with
  Spontaneously Broken Supersymmetry and their Effective Theories}},
  \href{http://dx.doi.org/10.1016/0550-3213(89)90048-5}{\emph{Nucl. Phys. B}
  {\bf 318} (1989) 75--105}.

\bibitem{Antoniadis:1990ew}
I.~Antoniadis, \emph{{A Possible new dimension at a few TeV}},
  \href{http://dx.doi.org/10.1016/0370-2693(90)90617-F}{\emph{Phys. Lett. B}
  {\bf 246} (1990) 377--384}.

\bibitem{Antoniadis:1991kh}
I.~Antoniadis and C.~Kounnas, \emph{{Superstring phase transition at high
  temperature}},
  \href{http://dx.doi.org/10.1016/0370-2693(91)90442-S}{\emph{Phys. Lett. B}
  {\bf 261} (1991) 369--378}.

\bibitem{Antoniadis:1998ki}
I.~Antoniadis, E.~Dudas and A.~Sagnotti, \emph{{Supersymmetry breaking, open
  strings and M theory}},
  \href{http://dx.doi.org/10.1016/S0550-3213(98)00806-2}{\emph{Nucl. Phys. B}
  {\bf 544} (1999) 469--502}, [\href{https://arxiv.org/abs/hep-th/9807011}{{\tt
  hep-th/9807011}}].

\bibitem{Antoniadis:1999xk}
I.~Antoniadis, E.~Dudas and A.~Sagnotti, \emph{{Brane supersymmetry breaking}},
  \href{http://dx.doi.org/10.1016/S0370-2693(99)01023-0}{\emph{Phys. Lett. B}
  {\bf 464} (1999) 38--45}, [\href{https://arxiv.org/abs/hep-th/9908023}{{\tt
  hep-th/9908023}}].

\bibitem{Ibanez:2017kvh}
L.~E. Ibanez, V.~Martin-Lozano and I.~Valenzuela, \emph{{Constraining Neutrino
  Masses, the Cosmological Constant and BSM Physics from the Weak Gravity
  Conjecture}}, \href{http://dx.doi.org/10.1007/JHEP11(2017)066}{\emph{JHEP}
  {\bf 11} (2017) 066}, [\href{https://arxiv.org/abs/1706.05392}{{\tt
  1706.05392}}].

\end{thebibliography}\endgroup

\end{document}